\documentclass[%
 reprint,
superscriptaddress,
 amsmath,amssymb,
 aps,
nolongbibliography
]{revtex4-2}
\usepackage{float}
\usepackage{graphicx}
\usepackage{dcolumn}
\usepackage{bm}
\usepackage[colorlinks=true, allcolors=blue]{hyperref}
\usepackage{xcolor}
\usepackage{bibentry}

\allowdisplaybreaks

\usepackage{physics}
\usepackage{amsmath}
\usepackage{amssymb}
\usepackage{mathrsfs}
\usepackage{subfigure}
\allowdisplaybreaks

\begin{document}


\title{Partial and full tunneling processes across potential barriers}

\author{Philip Caesar Flores}
\email{flores@mbi-berlin.de}
\affiliation{Max-Born-Institute, Max-Born Straße 2A, 12489 Berlin, Germany}
\author{Dean Alvin L. Pablico}
\email{dlpablico@up.edu.ph}
\author{Eric A. Galapon}%
\email{eagalapon@up.edu.ph}
\affiliation{%
	Theoretical Physics Group, National Institute of Physics \\ 
	University of the Philippines Diliman, 1101 Quezon City, Philippines
}%

%
%

\date{\today}

\begin{abstract}
We introduce the concept of partial-tunneling and full-tunneling processes to explain the seemingly contradictory non-zero and vanishing tunneling times often reported in the literature. Our analysis starts by considering the traversal time of a quantum particle through a potential barrier, including both above and below-barrier traversals, using the theory of time-of-arrival operators. We then show that there are three traversal processes corresponding to non-tunneling, full-tunneling, and partial tunneling. The distinction between the three depends on the support of the incident wavepacket’s energy distribution in relation to the shape of the barrier. Non-tunneling happens when the energy distribution of the quantum particle lies above the maximum of the potential barrier. Otherwise, full-tunneling process occurs when the energy distribution of the particle is below the minimum of the potential barrier. For this process, the obtained traversal time is interpreted as the tunneling time. Finally, the partial-tunneling process occurs when the energy distribution lies between the minimum and maximum of the potential barrier. This signifies that the quantum particle tunneled only through some portions of the potential barrier. We argue that the duration for a partial-tunneling process should not be interpreted as the tunneling time but instead as a partial traversal time to differentiate it from the full-tunneling process. We then show that a full-tunneling process is always instantaneous, while a partial-tunneling process takes a non-zero amount of time. We are then led to the hypothesis that experimentally measured non-zero and vanishing tunneling times correspond to partial and full-tunneling processes, respectively.
\end{abstract}

\maketitle

The time it takes for a quantum particle to tunnel through a potential barrier has always eluded physicists since the advent of quantum mechanics \cite{maccoll1932note,hartman1962tunneling}. A number of tunneling time definitions have been offered in the literature \cite{wigner1955lower,buttiker1982traversal,baz1966lifetime,rybachenko1967time,buttiker1983larmor,pollak1984new,smith1960lifetime,petersen2018instantaneous,sokolovski1987traversal,yamada2004unified,brouard1994systematic}, e.g., Wigner phase time \cite{wigner1955lower}, B\"{u}ttiker-Landauer time \cite{buttiker1982traversal}, Larmor time \cite{baz1966lifetime,rybachenko1967time,buttiker1983larmor}, Pollak-Miller time \cite{pollak1984new}, dwell time \cite{smith1960lifetime}; but a consensus on whether quantum tunneling is instantaneous or not is yet to be reached \cite{PhysRevA.101.052121,de2002time,winful2006tunneling,imafuku1997effects,jaworski1988time,leavens1989dwell,hauge1987transmission,hauge1989tunneling}. The development of ultraprecise techniques in strong-field physics has been expected to close the debate once and for all, but the existence of contradictory experimental results only further divided the physics community \cite{torlina2015interpreting,eckle2008attosecond,eckle2008attosecond2,pfeiffer2012attoclock,pfeiffer2013recent,sainadh2019attosecond,landsman2014ultrafast,camus2017experimental}. This diversity poses a challenge to theoretical treatments that only predict either zero or non-zero tunneling times, imploring a formalism that could accommodate both seemingly contradictory results. In this Letter, we offer such a formalism using the theory of time-of-arrival (TOA) operators \cite{galapon2018quantizations}. 

\begin{figure}[b!]
\centering
\includegraphics[width=0.45\textwidth]{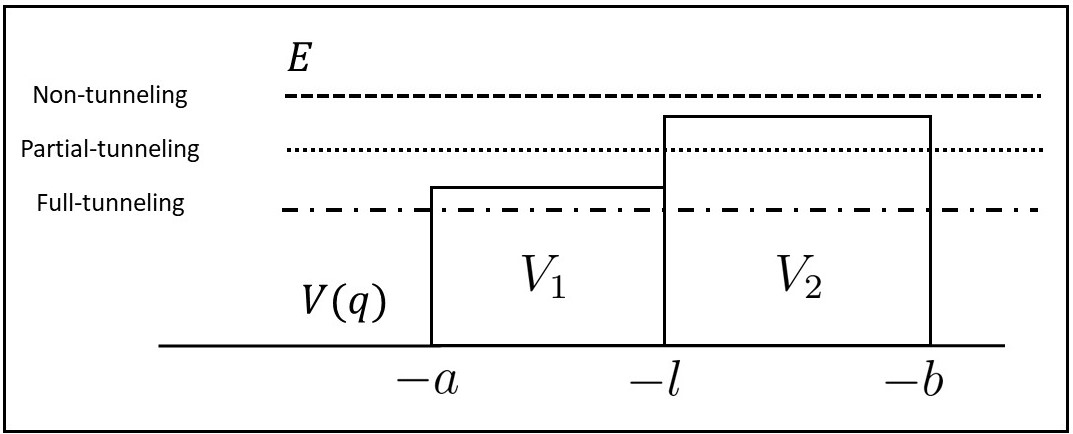}
\caption{Three traversal processes involving a quantum particle with incident energy $E$ traversing through a contiguous square barrier system.}
\label{fig:energycurve}
\end{figure}

We start our analysis by investigating the expected quantum traversal time across a contiguous barrier system, i.e., $V(q)=V_1$ for $-a<q<-l$ and $V(q)=V_2$ for $-l<q<-b$, where $V_1<V_2$. We define the traversal time as the amount of time a particle traverses through the barrier region, including both above-barrier and below-barrier traversals. We choose a contiguous potential barrier, instead of the usual square barrier, as it naturally showcases three possible traversal processes: (i) non-tunneling, (ii) full-tunneling, (iii) and partial-tunneling. The distinction between these three processes is shown in Fig. (\ref{fig:energycurve}). Non-tunneling or classical above-barrier traversal happens when the energy distribution $E$ of the quantum particle lies entirely above $V_2$. Meanwhile, full-tunneling process occurs when the energy distribution of the particle is entirely below $V_1$. The partial-tunneling process occurs when the energy distribution that lies between $V_1$ and $V_2$. This physically suggests that the quantum particle traversed above $V_1$ and tunneled through $V_2$, not the entire barrier region. The single square barrier can only describe the first two processes. We will show later that full-tunneling is always instantaneous while partial-tunneling and non-tunneling takes a non-zero amount of time. A double square barrier system has also been considered in Ref. \cite{sombillo2014quantum} but their analysis is focused on the generalized Hartman effect, which is different to the current paper's objectives. 

With clear distinctions between the three traversal processes, we now determine the corresponding quantum traversal time. We work under the assumption that there exists a TOA-operator $\hat{\mathrm{T}}$ corresponding to an arrival at some specific point in the configuration space for a given interaction potential $V(q)$. The traversal time can be extracted by determining the arrival time difference between two identical wave packets, the first one encounters a potential barrier while the other traverses freely without obstruction \cite{PhysRevLett.108.170402,sombillo2014quantum}.

\begin{figure}[t!]
\centering
\includegraphics[width=0.45\textwidth]{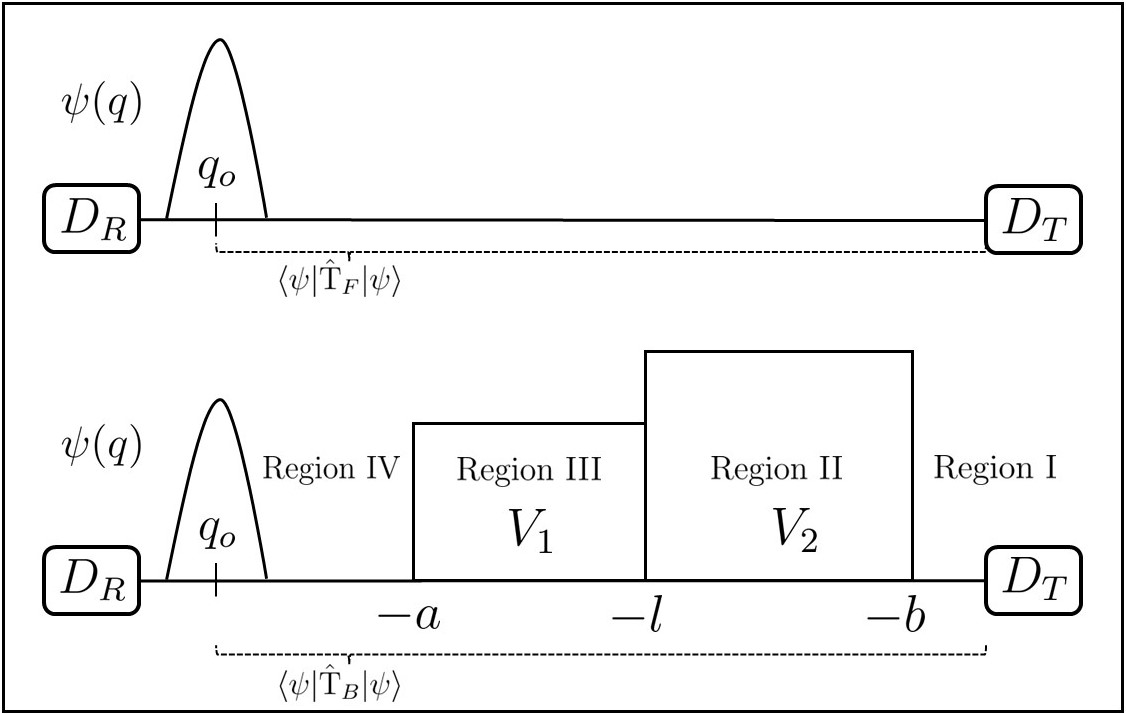}
\caption{Measurement scheme for the expected quantum traversal time.}
\label{fig:setup}
\end{figure}

Our measurement scheme follows the prescription of Refs. \cite{PhysRevLett.108.170402,sombillo2014quantum,pablico2020quantum,flores2023instantaneous,flores2023quantized} and shown in Fig. (\ref{fig:setup}). We place a detector $D_T$ at the far right of the barrier system to announce the arrival of the particle at the origin $q=0$. Likewise, a similar detector $D_R$ is placed at the far left of $D_T$. A localized wave packet $\psi(q)$ is launched in between the potential $V(q)$ and detector $D_R$ towards the origin at time $t=0.$ An arrival time is measured when $D_T$ clicks while no measurement is recorded when $D_R$ clicks. An average barrier TOA, $\bar{\tau}_B$, can be obtained when the same experiment is repeated over a large number of trials, with the same initial state for every measurement. A similar experiment is then performed in the absence of the potential barrier. The average free TOA, $\bar{\tau}_F$, is also measured with the detector $D_T$. The barrier traversal time is then extracted from the difference $\Delta\bar{\tau}=\bar{\tau}_F-\bar{\tau}_B$.

The above measurement scheme essentially coincides with the tunneling delay time used by Steinberg, Kwiat, and Chiao in their seminal single-photon tunneling time experiment \cite{steinberg1993measurement}. They employed a two-photon source in which pairs of photons are emitted simultaneously. One particle traverses a tunnel barrier while the other twin particle encounters no barrier. Their operational definition of tunneling time is extracted from the comparison between the TOA of the two conjugate particles. The difference, however, with our measurement scheme is that what they measured is $-\Delta\bar{\tau}$. 

Within the theory of TOA-operators, the measured average value $\bar{\tau}_B$ can be obtained as the expectation value of some barrier TOA-operator $\hat{\mathrm{T}}_B$ for a given state $\psi$, that is, $\bar{\tau}_B=\langle\psi| \hat{\mathrm{T}}_B|\psi\rangle$. Similarly, the average free time-of-arrival appears as $\bar{\tau}_F=\langle\psi|\hat{\mathrm{T}}_F|\psi\rangle,$ where $\hat{\mathrm{T}}_F$ is the corresponding free-particle TOA-operator. The arrival time difference then assumes the form
\begin{equation}
\Delta\bar{\tau} = \langle\psi|\hat{\mathrm{T}}_F|\psi\rangle - \langle\psi|\hat{\mathrm{T}}_B|\psi\rangle.
    \label{eq:toadiff}
\end{equation}
We highlight that $\Delta\bar{\tau}$ is not yet the barrier traversal time itself, but the latter can be obtained from it. 

Using the rigged Hilbert space formulation of quantum mechanics, a TOA-operator has the general form
\begin{equation}\label{eq:TOAdef}
(\hat{\mathrm{T}} \psi) (q) = \dfrac{\mu}{i\hbar}\int_{-\infty}^\infty dq'\, T(q,q') \,\mathrm{sgn}(q-q') \,\psi(q'),  
\end{equation}
in coordinate representation \cite{galapon2018quantizations}. The factor $\mathrm{sgn}z$ is the signum function, $\mu$ is the mass of the incident particle, and $ T(q,q')$ is referred to as the time kernel factor (TKF). The construction of the integral operator $\hat{\mathrm{T}}$ then translates to the construction of $T(q,q')$ for a given interaction potential $V(q)$. A closed-form expression for $T(q,q')$ can be obtained by performing Weyl-quantization on the classical TOA at the origin $\mathcal{T}_C(q,p)$ given by
\begin{equation}
\mathcal{T}_C(q,p)=-\mathrm{sgn}(p)\sqrt{\frac{\mu}{2}}\int_{0}^{q} \frac{dq'}{\sqrt{H(q,p)-V(q')}}.
\end{equation}
The quantization, however, should be restricted to the trajectories that pass through the arrival point and coincides with the first arrival of the particle \cite{galapon2018quantizations}. This gives the following Weyl-quantized TKF,
\begin{equation}\label{tkfweyl}
\tilde{T}(\eta,\zeta) = \dfrac{1}{2} \int_0^\eta ds \,{_0}F_1\left[ ; 1; \dfrac{\mu}{2 \hbar^2}(V(\eta)-V(s))\,\zeta^2\right],
\end{equation}
where, $T(q,q')=\tilde{T}(\eta,\zeta)$, $\eta=(q+q')/2$, $\zeta=q-q'$, and ${_0}F_1(;a;z)$ is a particular hypergeometric function \cite{galapon2018quantizations}. Equation (\ref{eq:TOAdef}), together with Eq. (\ref{tkfweyl}), define a TOA-operator that satisfies Hermiticity, time-reversal symmetry, and Dirac's correspondence between classical and quantum observables. 

Now, in the absence of a potential barrier, substitution of $V(q)=0$ into Eq. (\ref{tkfweyl}) yields the free TKF $\tilde{T}_F(\eta,\zeta)=\eta/2$ or equivalently $T_F(q,q')=(q+q')/4$. Substitution of $T_F(q,q')$ into Eq. (\ref{eq:TOAdef}) gives the free TOA-operator $\hat{\mathrm{T}}_F$ which is equivalent to $(\hat{\mathrm{T}}_{AB}\psi)(q)=\int dq' \langle q|\hat{\mathrm{T}}_{AB}|q'\rangle\psi(q')$, where $\hat{\mathrm{T}}_{AB}=-(\mu/2)(\hat{\mathrm{q}}\hat{\mathrm{p}}^{-1}+\hat{\mathrm{p}}^{-1}\hat{\mathrm{q}})$ is the well-known Aharonov-Bohm free time operator \cite{aharonov1961time}. Note that $\hat{\mathrm{T}}_F$ is canonically conjugate with the free Hamiltonian $\hat{\mathrm{H}}_F=\hat{\mathrm{p}}^2/2\mu.$

On the other hand, the barrier TOA-operator $\hat{\mathrm{T}}_B$ is constructed by solving for the barrier TKF $T_B(q,q')=\tilde{T}_B(\eta,\zeta)$. This is done by dividing the $s$ integral in Eq. (\ref{tkfweyl}) into four non-overlapping regions separated by the edges of the two distinct barriers. The barrier TKF $\tilde{T}_B(\eta,\zeta)$ then have four pieces corresponding to the four regions described in Fig. (\ref{fig:setup}) and are given by
\begin{subequations}\label{timekernel}
\begin{align}
\tilde{T}_{B,I}(\eta,\zeta)=&\frac{\eta}{2}, \\
\tilde{T}_{B,II}(\eta,\zeta)=&\frac{\eta+b}{2}-\frac{w_1}{2}I_0(\kappa_1|\zeta|), \\
\tilde{T}_{B,III}(\eta,\zeta)=&\frac{\eta+b+w_1}{2}-\frac{b}{2}I_0(\kappa_2|\zeta|)\\&-\frac{w_1}{2}{_0F_1}\left(;1;\frac{\kappa^2_{21}}{4}\zeta^2\right), \\
\tilde{T}_{B,IV}(\eta,\zeta)=&\frac{\eta+L}{2}-\sum_{n=1,2}\frac{w_n}{2}J_0(\kappa_n |\zeta|),
\end{align}
\end{subequations}
where $\kappa_n=\sqrt{2\mu V_n}/\hbar$ for $n=1,2$, $\kappa^2_{2,1}=2\mu(V_2-V_1)/\hbar^2,$ and $w_n$ are the widths of each barrier \cite{sombillo2014quantum}. The above results have been derived using the identities $_{0}F_1(;1,x)=I_0(2\sqrt{x})$ for $x>0$ and $_{0}F_1(;1,x)=J_0(2\sqrt{x})$ for $x<0$, where $I_0(z)$ and $J_0(z)$ are specific Bessel functions. Substitution of Eq. (\ref{timekernel}) into Eq. (\ref{eq:TOAdef}) gives the barrier TOA-operator $\hat{\mathrm{T}}_B$. It can be shown that $\hat{\mathrm{T}}_B$ with the TKF $\tilde{T}_{B,r}(\eta,\zeta)$ for $r=I,II,III,IV$ is canonically conjugate with the Hamiltonians in the respective regions, $r$ \cite{galapon2004shouldn}.

Having constructed the operators $\hat{\mathrm{T}}_B$ and $\hat{\mathrm{T}}_F$, we now evaluate the arrival time difference $\Delta\bar{\tau}$ in Eq. (\ref{eq:toadiff}). We assume an incident wave packet of the form $\psi(q) = e^{ik_oq/\hbar}\varphi(q)$ centered at $q=q_o$ with a mean momentum expectation value $p_o=\hbar k_o$. We also impose the support of $\varphi(q)$ to lie entirely to the left of the barrier system. The latter assumption suggests that there is a zero probability that the quantum particle is already within the barrier region or at the transmission side at the initial time $t=0$. Under this condition, our barrier TOA-operator  $\hat{\mathrm{T}}_B$ depends only on the piece $\tilde{T}_{B,IV}(\eta,\zeta)$. 

In the $(\eta,\zeta)$ coordinates, the arrival time difference can be rewritten in the form
\begin{equation}\label{complextoadiff}
\Delta\bar{\tau}=-\frac{2\mu}{\hbar}\,\mathrm{Im}\,\int_{0}^{\infty} d\zeta\,\Phi(\zeta)\, e^{i k _0\zeta}\,\left[ \tilde{T}_F(\eta)-\tilde{T}_{B,IV}(\eta,\zeta) \right],
\end{equation}
where, $\Phi(\zeta)=\int_{-\infty}^{\infty} d\eta\,\bar{\varphi}(\eta-\zeta/2)\,\varphi(\eta+\zeta/2)$, 
and $\mathrm{Im}(z)$ denotes the imaginary part of the integral. Evaluation of Eq. (\ref{complextoadiff}) leads to
\begin{subequations}
\begin{align}
\Delta\bar{\tau}=&\dfrac{L}{v_0} \mathrm{Im}\, Q^* \,-\,\sum_{n=1}^2\,\dfrac{w_n}{v_0}\,\mathrm{Im}\,R_n^* \label{toadiff2}, \\
Q^*=&ko \int_0^\infty d\zeta \Phi(\zeta)\, e^{i k _0\zeta}, \\
R_n^*=&ko \int_0^\infty d\zeta \Phi(\zeta)\, J_0(\kappa_n \zeta)\,e^{i k _0\zeta}.
\end{align}
\end{subequations}
We determine the physical significance of $\Delta\bar{\tau}$ by taking its classical limit. This is done by considering the high energy limit $k_0 \to \infty$ for fixed $\kappa_n.$ One then finds the factor $\mathrm{Im}\, Q^* \sim 1$ while $\mathrm{Im}\,R_n^* \sim v_0/v_n$, where $v_n$ is the particle's speed on top of the barrier with width $w_n$. Since $\mathrm{Im}\,R_n^*$ depends on the ratio between the speeds $v_0$ and $v_n$, we can interpret it as the index of refraction of the $n$th barrier. Hence, we find the relation $\Delta\bar{\tau} \sim L/v_0-\,\sum_{n=1}^2\,w_n/v_n.$ The first term is simply the free classical TOA across the region of length $L$. On the other hand, the second term is identified as the classical barrier traversal time.

The above classical limits suggest that the expected quantum traversal time across the barrier region can be extracted from the second term of Eq. (\ref{toadiff2}). Evaluation of $R_n^*$ using the Fourier transform of the full incident wave function $\psi(q)$, i.e., $\tilde{\psi}(k)=(2\pi)^{-1/2}\int_{-\infty}^{\infty }e^{-ikq} \psi(q)$, the second term leads to
\begin{equation}
\bar{\tau}_{B}=\sum_{n=1}^2 \dfrac{w_n}{\nu_0}  \int_{\kappa_n}^{\infty} dk \dfrac{|\tilde{\psi}(k)|^2-|\tilde{\psi}(-k)|^2}{\sqrt{k^2 - \kappa_n^2}}.
\label{barriertrav}
\end{equation}
Equation (\ref{barriertrav}) clearly shows the contribution of the positive and negative momentum components of the incident wave function  $\tilde{\psi}(k)$. Within our framework, $\bar{\tau}_{B}$ can be interpreted as the dwell time in the barrier region which is the total average time that our incident particle spends in the barrier, regardless of transmission (where the positive components dominate) or reflection. 

The measurable quantum traversal time $\bar{\tau}_{trav}$ at the transmission channel then comes from the positive momentum components. Hence, we can now define the barrier traversal time as
\begin{equation}\label{trav}
\bar{\tau}_{trav}=\,\sum_{n=1}^2\,\dfrac{\mu \,w_n}{\hbar \,k_0}\,\int_{\kappa_n}^{\infty} dk\,\dfrac{|\tilde{\psi}(k)|^2}{\sqrt{k^2 - \kappa_n^2}}.
\end{equation}
To extract the time durations for the three tunneling processes, we rewrite Eq. \eqref{trav} as
\begin{subequations}
\begin{align}
\bar{\tau}_{\text{trav}}=&\dfrac{L}{\nu_o}(R_{\text{part}} + R_{\text{non}}) \label{eq:trav}, \\
R_{\text{part}} =& \dfrac{w_1 k_o}{L} \int_{\kappa_1}^{\kappa_2} dk \dfrac{|\tilde{\psi}(k)|^2}{\sqrt{k^2-\kappa_1^2}} \label{eq:Rpart},\\
R_{\text{non}} =& \dfrac{k_o}{L} \sum_{n=1}^2 w_n\int_{\kappa_2}^\infty dk \dfrac{|\tilde{\psi}(k)|^2}{\sqrt{k^2-\kappa_n^2}}. \label{eq:Rnon}
\end{align}
\end{subequations}
Both Eqs. (\ref{trav}) and (\ref{eq:trav}) clearly suggest that only the energy components satisfying $k>\kappa_1$ contribute to any measurable quantum traversal time, in keeping with the results of Refs. \cite{PhysRevLett.108.170402,sombillo2014quantum}. In addition, we can extract three specific traversal time regimes from  Eq. (\ref{eq:trav}) depending on the support of the distribution $|\tilde{\psi}(k)|^2$. These regimes coincide with our definition of (i) non-tunneling or classical above-barrier traversal, (ii) full-tunneling, and (iii) partial-tunneling processes. 

The non-tunneling regime occurs when the distribution $|\tilde{\psi}(k)|^2$ has a compact support in $k>\kappa_2$ such that the full energy distribution of the incident wave packet lies above the two potential barrier heights. The quantity $\bar{\tau}_{\text{non}}=(L/\nu_o)R_{\text{non}}$ from Eq. \eqref{eq:Rnon} then gives the expected above-barrier quantum traversal time and can be rewritten as
\begin{equation}
\bar{\tau}_{\text{non}}=\,\sum_{n=1}^2\,\dfrac{\mu w_n}{\hbar k_0}\,\int_{\kappa_2}^{\infty} dk\,|\tilde{\psi}(k)|^2 \,\tau(k),
\end{equation}
where $\tau(k)=\hbar\sqrt{k^2 - \kappa_n^2}/\mu$ are just the classical traversal times on top of entire barrier system. 

The full-tunneling regime occurs when the support of $|\tilde{\psi}(k)|^2$ lies below $\kappa_1$, so that Eq. (\ref{trav}), or equivalently Eq. (\ref{eq:trav}), results to a vanishing traversal time. Since all the energy components are below the barrier heights, the vanishing traversal time is interpreted as instantaneous tunneling time, that is,
\begin{equation}\label{fullttime}
\bar{\tau}_{\text{tun}} =0.
\end{equation}
Last, the partial-tunneling regime occurs when the momentum distribution $|\tilde{\psi}(k)|^2$ has components that lie between $\kappa_1<k<\kappa_2$. These components that lie between $\kappa_1<k<\kappa_2$ corresponds to a particle that traverses above $V_1$ and tunnels through $V_2$. Thus, the particle did not tunnel through the entire barrier system, hence, the name partial-tunneling. The quantity $\bar{\tau}_{\text{part}} = (L/\nu_o)R_{\text{part}}$ is now interpreted as a ``partial-traversal time'' since Eq. \eqref{eq:Rpart} indicates that the measured value originates from the momentum components that traversed above the barrier $V_1$.  

We wish to highlight that the resolution of the quantum tunneling time problem starts when everyone agrees that the term \textit{tunneling time} be used to describe a full-tunneling process. When an incident particle only partially tunnels through a barrier region, one should avoid using the term tunneling time. For our case, we used the term \textit{partial-traversal time}. Within our framework, we conclude that quantum tunneling, whenever it happens, is always instantaneous. 

The above results hold in general. We can model an arbitrary potential barrier $V(q)$ as a system of composite square barriers with varying heights and widths, i.e., $V(q)=\sum_{n=1}^\infty V_n$ each having a width $w_n$. In the continuous limit, $w_n\rightarrow0$, the barrier traversal time is
\begin{subequations}
\begin{align}
\bar{\tau}_{\text{trav}} =& \frac{L}{\nu_o} (R_{\text{part}} + R_{\text{non}}), \label{eq:trav_gen} \\
R_{\text{part}} =&  \dfrac{k_o}{L} \int_{b}^a dx   \int_{\kappa(x)}^{\kappa_{\text{max}}} dk  \dfrac{|\tilde{\psi}(k)|^2}{\sqrt{k^2-\kappa(x)^2}} \label{eq:Rpart_gen}, \\
R_{\text{non}} =& \dfrac{k_o}{L} \int_{b}^a dx \int_{\kappa_{\text{max}}}^\infty dk  \dfrac{|\tilde{\psi}(k)|^2}{\sqrt{k^2-\kappa(x)^2}},\label{eq:Rnon_gen}
\end{align}
\end{subequations}
where $\kappa_{\text{max}} = \sqrt{2\mu V_{\text{max}} }/\hbar$ indicates the maximum value of the the barrier height and $\kappa(x) = \sqrt{2\mu V(x)/\hbar^2}$. 

Notice the distinction between traversal time and tunneling time. For the general case when a wave packet has both above, below, and in-between barrier energy components, the traversal time through a potential barrier is the sum of the vanishing tunneling time $\bar{\tau}_{\text{tun}}$, non-zero partial traversal time $\bar{\tau}_{\text{part}}$, and above-barrier traversal time $\bar{\tau}_{\text{non}}$, i.e.,
\begin{equation}
    \bar{\tau}_{\text{trav}} = \bar{\tau}_{\text{tun}} + \bar{\tau}_{\text{part}} + \bar{\tau}_{\text{non}} = \bar{\tau}_{\text{part}} + \bar{\tau}_{\text{non}}.
\end{equation}
If one does not properly differentiate partial and full-tunneling processes, the contribution of the partial traversal time may be mistakenly identified as the tunneling time especially if one only considers tunneling with respect to $\kappa_{\text{max}}$.

For a square potential barrier system, we can simply take $V(x)=V_0$ so that $\bar{\tau}_{\text{part}}= \bar{\tau}_{\text{tun}} =0$. Quantum tunneling for this case is only described by a full-tunneling process, and we recover the predictions of Ref. \cite{PhysRevLett.108.170402} that only above barrier energy components contribute to the barrier traversal time. For smooth barriers with compact support, Eq. \eqref{eq:Rpart_gen} indicates that the partial traversal time $\bar{\tau}_{\text{part}}=(L/\nu_o)R_{\text{part}}$ will always be non-zero since the $\kappa_{\text{min}}=0$, provided that a segment of the support of $\tilde{\psi}(k)$ is below $\kappa_{\text{max}}$. Thus, quantum tunneling for this case is only described by a partial-tunneling process.

The main advantage of our treatment in comparison with other tunneling time definitions is its simplicity and generality. Specifically, other tunneling time definitions such as the B\"{u}ttiker-Landauer time, Larmor time, and Pollak-Miller time involves calculating the transmission amplitude for propagating through the barrier, which will require solving the Sch\"{o}dinger equation. However, our treatment only requires information on the incident wavepacket and the interaction potential, which allows it to be applicable in any system without the need of any further calculations. Of course, our current analysis is anchored on the assumption that the incident wavepacket does not initially `leak' into the barrier. In addition, our treatment encompasses both non-zero and zero traversal times depending on the initial state of the particle and the shape of the potential barrier system. This is not obtained in other approaches so it makes sense why they cannot explain the seemingly contradictory reports in tunneling time experiments.

Following our results, we conclude that the non-zero tunneling time reported in Ref. \cite{steinberg1993measurement} is due to a partial-tunneling process. Specifically, the tunnel barrier used was a multilayer dielectric mirror that has an $(HL)^5H$ structure, where $H$ represents titanium oxide with a refractive index of $n_H=2.22$ while $L$ represents fused silica with a refractive index of $n_L=1.41$. This setup is an optical analogue of a system of contiguous square barriers with heights $V_H$ and $V_L$, where $V_L<V_H$, and can be considered as a potential barrier with jump discontinuities. Since there was no way that the momentum distribution of the incident photon can be controlled such that all the momentum components are below $V_L$, then the photon exhibits \textit{partial tunneling} resulting to a non-zero partial traversal time. 

\begin{figure}[t!]
\centering
\includegraphics[width=0.45\textwidth]{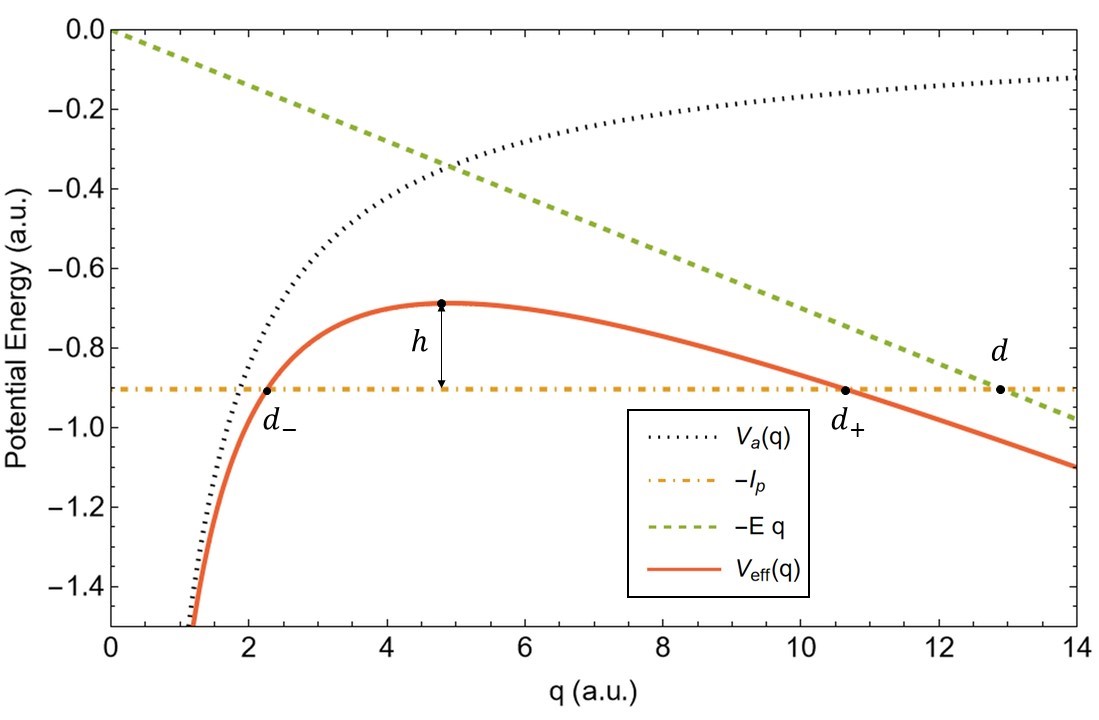}
\caption{Potential curves for a Helium atom in the presence of an external electric field $E$ with $Z_{eff}=1.6875$, $I_p=0.90357 \mathrm{au}$. In strong-field physics, an electron with bound state energy $-I_p$ tunnels through the an effective potential barrier $V_{eff}=-Z_{eff}/q-Eq$ shown in the red line. The above model is based on the analysis of Ref. \cite{Kullie2020}.}
\label{fig:potentialcurves}
\end{figure}

We also argue that $\bar{\tau}_{\text{part}}$ may explain any upper bound in the tunneling time of attoclock experiments which reported instantaneous tunneling time \cite{eckle2008attosecond2,pfeiffer2012attoclock,pfeiffer2013recent,torlina2015interpreting,sainadh2019attosecond}. In particular, theoretical modelling of quantum tunneling in strong field-physics based on Simpleman's model considers an electron which starts from a bound state and tunnels through the barrier to become a continuum state. The atomic Coulomb potential $V_a(q)=-Z_{eff}/q$ is distorted by an external electric field $E$ to create an effective potential barrier $V_{eff}(q)=-Z_{eff}/q-E q$ as shown in Fig. (\ref{fig:potentialcurves}), where $Z_{eff} $ is the effective nuclear charge \cite{Kullie2020}. Semiclassical analysis of tunneling time considers the electron traversing through the barrier $V_{eff}(q)$ of length $d$ with point $d$ being considered as the \textit{classical} tunnel exit. The corresponding tunneling time is called the Keldysh time. 

It has been argued that the Keldysh time may substantially exceed the actual tunneling time \cite{Sainadh2020}. In our formalism, this process is exactly a partial-tunneling process since the particle has an above-barrier energy component in the region between the turning point $d_+$ and classical exit $d$. Assuming there is a zero probability that the particle is already in the barrier region, the non-zero Keldysh time corresponds to our non-vanishing partial-traversal time and the reason why the former exceeds the actual tunneling time is due to the contribution of the above-barrier energy components of the incident particle. Hence, the Keldysh time should not be interpreted as the tunneling time. Another way to show a substantial non-vanishing traversal time is when the support of the particle's initial state extends into the barrier region which follows from the result of Ref. \cite{sombillo2014quantum}. 

On the other hand, quantum tunneling based on Kullie's model considers tunneling where a particle enters the barrier region at point $d_-$ and exits at point $d_+$ \cite{Kullie2015}. It is found that tunneling time decreases as a function of increasing field strength $E$ \cite{Kullie2015,Sainadh2020}. In our framework, this can be explained by the decrease in the contribution of the above-barrier energy components of the incident particle. For sufficiently strong electric field, all energy components are exactly below the effective potential barrier so that we get an instantaneous quantum tunneling.

It is often mentioned that the tunneling time using attoclock ionization technique gives seemingly contradictory results. Some experiments suggest an instantaneous quantum tunneling while some find the opposite result. One may argue that our model does not exactly coincide with attoclock experiments. However this is not entirely true since the character of these experiments still reduce to our current model. That is, a quantum particle travels through a potential barrier and is detected at the other side. The time it takes to traverse the barrier is then measured. Following our results, we are led to the hypothesis that experimentally measured non-zero tunneling times correspond to partial-tunneling processes while vanishing tunneling times correspond to full-tunneling processes.

\bibliography{tunnelregimes-letter.bib}

\begin{thebibliography}{41}%
\makeatletter
\providecommand \@ifxundefined [1]{%
 \@ifx{#1\undefined}
}%
\providecommand \@ifnum [1]{%
 \ifnum #1\expandafter \@firstoftwo
 \else \expandafter \@secondoftwo
 \fi
}%
\providecommand \@ifx [1]{%
 \ifx #1\expandafter \@firstoftwo
 \else \expandafter \@secondoftwo
 \fi
}%
\providecommand \natexlab [1]{#1}%
\providecommand \enquote  [1]{``#1''}%
\providecommand \bibnamefont  [1]{#1}%
\providecommand \bibfnamefont [1]{#1}%
\providecommand \citenamefont [1]{#1}%
\providecommand \href@noop [0]{\@secondoftwo}%
\providecommand \href [0]{\begingroup \@sanitize@url \@href}%
\providecommand \@href[1]{\@@startlink{#1}\@@href}%
\providecommand \@@href[1]{\endgroup#1\@@endlink}%
\providecommand \@sanitize@url [0]{\catcode `\\12\catcode `\$12\catcode
  `\&12\catcode `\#12\catcode `\^12\catcode `\_12\catcode `\%12\relax}%
\providecommand \@@startlink[1]{}%
\providecommand \@@endlink[0]{}%
\providecommand \url  [0]{\begingroup\@sanitize@url \@url }%
\providecommand \@url [1]{\endgroup\@href {#1}{\urlprefix }}%
\providecommand \urlprefix  [0]{URL }%
\providecommand \Eprint [0]{\href }%
\providecommand \doibase [0]{https://doi.org/}%
\providecommand \selectlanguage [0]{\@gobble}%
\providecommand \bibinfo  [0]{\@secondoftwo}%
\providecommand \bibfield  [0]{\@secondoftwo}%
\providecommand \translation [1]{[#1]}%
\providecommand \BibitemOpen [0]{}%
\providecommand \bibitemStop [0]{}%
\providecommand \bibitemNoStop [0]{.\EOS\space}%
\providecommand \EOS [0]{\spacefactor3000\relax}%
\providecommand \BibitemShut  [1]{\csname bibitem#1\endcsname}%
\let\auto@bib@innerbib\@empty
\bibitem [{\citenamefont {MacColl}(1932)}]{maccoll1932note}%
  \BibitemOpen
  \bibfield  {author} {\bibinfo {author} {\bibfnamefont {L.}~\bibnamefont
  {MacColl}},\ }\href@noop {} {\bibfield  {journal} {\bibinfo  {journal}
  {Physical Review}\ }\textbf {\bibinfo {volume} {40}},\ \bibinfo {pages} {621}
  (\bibinfo {year} {1932})}\BibitemShut {NoStop}%
\bibitem [{\citenamefont {Hartman}(1962)}]{hartman1962tunneling}%
  \BibitemOpen
  \bibfield  {author} {\bibinfo {author} {\bibfnamefont {T.~E.}\ \bibnamefont
  {Hartman}},\ }\href@noop {} {\bibfield  {journal} {\bibinfo  {journal}
  {Journal of Applied Physics}\ }\textbf {\bibinfo {volume} {33}},\ \bibinfo
  {pages} {3427} (\bibinfo {year} {1962})}\BibitemShut {NoStop}%
\bibitem [{\citenamefont {Wigner}(1955)}]{wigner1955lower}%
  \BibitemOpen
  \bibfield  {author} {\bibinfo {author} {\bibfnamefont {E.~P.}\ \bibnamefont
  {Wigner}},\ }\href@noop {} {\bibfield  {journal} {\bibinfo  {journal}
  {Physical Review}\ }\textbf {\bibinfo {volume} {98}},\ \bibinfo {pages} {145}
  (\bibinfo {year} {1955})}\BibitemShut {NoStop}%
\bibitem [{\citenamefont {B{\"u}ttiker}\ and\ \citenamefont
  {Landauer}(1982)}]{buttiker1982traversal}%
  \BibitemOpen
  \bibfield  {author} {\bibinfo {author} {\bibfnamefont {M.}~\bibnamefont
  {B{\"u}ttiker}}\ and\ \bibinfo {author} {\bibfnamefont {R.}~\bibnamefont
  {Landauer}},\ }\href@noop {} {\bibfield  {journal} {\bibinfo  {journal}
  {Physical Review Letters}\ }\textbf {\bibinfo {volume} {49}},\ \bibinfo
  {pages} {1739} (\bibinfo {year} {1982})}\BibitemShut {NoStop}%
\bibitem [{\citenamefont {Baz}(1966)}]{baz1966lifetime}%
  \BibitemOpen
  \bibfield  {author} {\bibinfo {author} {\bibfnamefont {A.}~\bibnamefont
  {Baz}},\ }\href@noop {} {\bibfield  {journal} {\bibinfo  {journal} {Yadern.
  Fiz.}\ }\textbf {\bibinfo {volume} {4}} (\bibinfo {year} {1966})}\BibitemShut
  {NoStop}%
\bibitem [{\citenamefont {Rybachenko}(1967)}]{rybachenko1967time}%
  \BibitemOpen
  \bibfield  {author} {\bibinfo {author} {\bibfnamefont {V.}~\bibnamefont
  {Rybachenko}},\ }\href@noop {} {\bibfield  {journal} {\bibinfo  {journal}
  {Sov. J. Nucl. Phys.}\ }\textbf {\bibinfo {volume} {5}},\ \bibinfo {pages}
  {635} (\bibinfo {year} {1967})}\BibitemShut {NoStop}%
\bibitem [{\citenamefont {B{\"u}ttiker}(1983)}]{buttiker1983larmor}%
  \BibitemOpen
  \bibfield  {author} {\bibinfo {author} {\bibfnamefont {M.}~\bibnamefont
  {B{\"u}ttiker}},\ }\href@noop {} {\bibfield  {journal} {\bibinfo  {journal}
  {Physical Review B}\ }\textbf {\bibinfo {volume} {27}},\ \bibinfo {pages}
  {6178} (\bibinfo {year} {1983})}\BibitemShut {NoStop}%
\bibitem [{\citenamefont {Pollak}\ and\ \citenamefont
  {Miller}(1984)}]{pollak1984new}%
  \BibitemOpen
  \bibfield  {author} {\bibinfo {author} {\bibfnamefont {E.}~\bibnamefont
  {Pollak}}\ and\ \bibinfo {author} {\bibfnamefont {W.~H.}\ \bibnamefont
  {Miller}},\ }\href@noop {} {\bibfield  {journal} {\bibinfo  {journal}
  {Physical review letters}\ }\textbf {\bibinfo {volume} {53}},\ \bibinfo
  {pages} {115} (\bibinfo {year} {1984})}\BibitemShut {NoStop}%
\bibitem [{\citenamefont {Smith}(1960)}]{smith1960lifetime}%
  \BibitemOpen
  \bibfield  {author} {\bibinfo {author} {\bibfnamefont {F.~T.}\ \bibnamefont
  {Smith}},\ }\href@noop {} {\bibfield  {journal} {\bibinfo  {journal}
  {Physical Review}\ }\textbf {\bibinfo {volume} {118}},\ \bibinfo {pages}
  {349} (\bibinfo {year} {1960})}\BibitemShut {NoStop}%
\bibitem [{\citenamefont {Petersen}\ and\ \citenamefont
  {Pollak}(2018)}]{petersen2018instantaneous}%
  \BibitemOpen
  \bibfield  {author} {\bibinfo {author} {\bibfnamefont {J.}~\bibnamefont
  {Petersen}}\ and\ \bibinfo {author} {\bibfnamefont {E.}~\bibnamefont
  {Pollak}},\ }\href@noop {} {\bibfield  {journal} {\bibinfo  {journal} {The
  Journal of Physical Chemistry A}\ }\textbf {\bibinfo {volume} {122}},\
  \bibinfo {pages} {3563} (\bibinfo {year} {2018})}\BibitemShut {NoStop}%
\bibitem [{\citenamefont {Sokolovski}\ and\ \citenamefont
  {Baskin}(1987)}]{sokolovski1987traversal}%
  \BibitemOpen
  \bibfield  {author} {\bibinfo {author} {\bibfnamefont {D.}~\bibnamefont
  {Sokolovski}}\ and\ \bibinfo {author} {\bibfnamefont {L.}~\bibnamefont
  {Baskin}},\ }\href@noop {} {\bibfield  {journal} {\bibinfo  {journal}
  {Physical Review A}\ }\textbf {\bibinfo {volume} {36}},\ \bibinfo {pages}
  {4604} (\bibinfo {year} {1987})}\BibitemShut {NoStop}%
\bibitem [{\citenamefont {Yamada}(2004)}]{yamada2004unified}%
  \BibitemOpen
  \bibfield  {author} {\bibinfo {author} {\bibfnamefont {N.}~\bibnamefont
  {Yamada}},\ }\href@noop {} {\bibfield  {journal} {\bibinfo  {journal}
  {Physical review letters}\ }\textbf {\bibinfo {volume} {93}},\ \bibinfo
  {pages} {170401} (\bibinfo {year} {2004})}\BibitemShut {NoStop}%
\bibitem [{\citenamefont {Brouard}\ \emph {et~al.}(1994)\citenamefont
  {Brouard}, \citenamefont {Sala},\ and\ \citenamefont
  {Muga}}]{brouard1994systematic}%
  \BibitemOpen
  \bibfield  {author} {\bibinfo {author} {\bibfnamefont {S.}~\bibnamefont
  {Brouard}}, \bibinfo {author} {\bibfnamefont {R.}~\bibnamefont {Sala}},\ and\
  \bibinfo {author} {\bibfnamefont {J.}~\bibnamefont {Muga}},\ }\href@noop {}
  {\bibfield  {journal} {\bibinfo  {journal} {Physical Review A}\ }\textbf
  {\bibinfo {volume} {49}},\ \bibinfo {pages} {4312} (\bibinfo {year}
  {1994})}\BibitemShut {NoStop}%
\bibitem [{\citenamefont {Yusofsani}\ and\ \citenamefont
  {Kolesik}(2020)}]{PhysRevA.101.052121}%
  \BibitemOpen
  \bibfield  {author} {\bibinfo {author} {\bibfnamefont {S.}~\bibnamefont
  {Yusofsani}}\ and\ \bibinfo {author} {\bibfnamefont {M.}~\bibnamefont
  {Kolesik}},\ }\href {https://doi.org/10.1103/PhysRevA.101.052121} {\bibfield
  {journal} {\bibinfo  {journal} {Phys. Rev. A}\ }\textbf {\bibinfo {volume}
  {101}},\ \bibinfo {pages} {052121} (\bibinfo {year} {2020})}\BibitemShut
  {NoStop}%
\bibitem [{\citenamefont {de~Carvalho}\ and\ \citenamefont
  {Nussenzveig}(2002)}]{de2002time}%
  \BibitemOpen
  \bibfield  {author} {\bibinfo {author} {\bibfnamefont {C.~A.}\ \bibnamefont
  {de~Carvalho}}\ and\ \bibinfo {author} {\bibfnamefont {H.~M.}\ \bibnamefont
  {Nussenzveig}},\ }\href@noop {} {\bibfield  {journal} {\bibinfo  {journal}
  {Physics Reports}\ }\textbf {\bibinfo {volume} {364}},\ \bibinfo {pages} {83}
  (\bibinfo {year} {2002})}\BibitemShut {NoStop}%
\bibitem [{\citenamefont {Winful}(2006)}]{winful2006tunneling}%
  \BibitemOpen
  \bibfield  {author} {\bibinfo {author} {\bibfnamefont {H.~G.}\ \bibnamefont
  {Winful}},\ }\href@noop {} {\bibfield  {journal} {\bibinfo  {journal}
  {Physics Reports}\ }\textbf {\bibinfo {volume} {436}},\ \bibinfo {pages} {1}
  (\bibinfo {year} {2006})}\BibitemShut {NoStop}%
\bibitem [{\citenamefont {Imafuku}\ \emph {et~al.}(1997)\citenamefont
  {Imafuku}, \citenamefont {Ohba},\ and\ \citenamefont
  {Yamanaka}}]{imafuku1997effects}%
  \BibitemOpen
  \bibfield  {author} {\bibinfo {author} {\bibfnamefont {K.}~\bibnamefont
  {Imafuku}}, \bibinfo {author} {\bibfnamefont {I.}~\bibnamefont {Ohba}},\ and\
  \bibinfo {author} {\bibfnamefont {Y.}~\bibnamefont {Yamanaka}},\ }\href@noop
  {} {\bibfield  {journal} {\bibinfo  {journal} {Physical Review A}\ }\textbf
  {\bibinfo {volume} {56}},\ \bibinfo {pages} {1142} (\bibinfo {year}
  {1997})}\BibitemShut {NoStop}%
\bibitem [{\citenamefont {Jaworski}\ and\ \citenamefont
  {Wardlaw}(1988)}]{jaworski1988time}%
  \BibitemOpen
  \bibfield  {author} {\bibinfo {author} {\bibfnamefont {W.}~\bibnamefont
  {Jaworski}}\ and\ \bibinfo {author} {\bibfnamefont {D.~M.}\ \bibnamefont
  {Wardlaw}},\ }\href@noop {} {\bibfield  {journal} {\bibinfo  {journal}
  {Physical Review A}\ }\textbf {\bibinfo {volume} {38}},\ \bibinfo {pages}
  {5404} (\bibinfo {year} {1988})}\BibitemShut {NoStop}%
\bibitem [{\citenamefont {Leavens}\ and\ \citenamefont
  {Aers}(1989)}]{leavens1989dwell}%
  \BibitemOpen
  \bibfield  {author} {\bibinfo {author} {\bibfnamefont {C.}~\bibnamefont
  {Leavens}}\ and\ \bibinfo {author} {\bibfnamefont {G.}~\bibnamefont {Aers}},\
  }\href@noop {} {\bibfield  {journal} {\bibinfo  {journal} {Physical Review
  B}\ }\textbf {\bibinfo {volume} {39}},\ \bibinfo {pages} {1202} (\bibinfo
  {year} {1989})}\BibitemShut {NoStop}%
\bibitem [{\citenamefont {Hauge}\ \emph {et~al.}(1987)\citenamefont {Hauge},
  \citenamefont {Falck},\ and\ \citenamefont
  {Fjeldly}}]{hauge1987transmission}%
  \BibitemOpen
  \bibfield  {author} {\bibinfo {author} {\bibfnamefont {E.}~\bibnamefont
  {Hauge}}, \bibinfo {author} {\bibfnamefont {J.}~\bibnamefont {Falck}},\ and\
  \bibinfo {author} {\bibfnamefont {T.}~\bibnamefont {Fjeldly}},\ }\href@noop
  {} {\bibfield  {journal} {\bibinfo  {journal} {Physical Review B}\ }\textbf
  {\bibinfo {volume} {36}},\ \bibinfo {pages} {4203} (\bibinfo {year}
  {1987})}\BibitemShut {NoStop}%
\bibitem [{\citenamefont {Hauge}\ and\ \citenamefont
  {St{\o}vneng}(1989)}]{hauge1989tunneling}%
  \BibitemOpen
  \bibfield  {author} {\bibinfo {author} {\bibfnamefont {E.}~\bibnamefont
  {Hauge}}\ and\ \bibinfo {author} {\bibfnamefont {J.}~\bibnamefont
  {St{\o}vneng}},\ }\href@noop {} {\bibfield  {journal} {\bibinfo  {journal}
  {Reviews of Modern Physics}\ }\textbf {\bibinfo {volume} {61}},\ \bibinfo
  {pages} {917} (\bibinfo {year} {1989})}\BibitemShut {NoStop}%
\bibitem [{\citenamefont {Torlina}\ \emph {et~al.}(2015)\citenamefont
  {Torlina}, \citenamefont {Morales}, \citenamefont {Kaushal}, \citenamefont
  {Ivanov}, \citenamefont {Kheifets}, \citenamefont {Zielinski}, \citenamefont
  {Scrinzi}, \citenamefont {Muller}, \citenamefont {Sukiasyan}, \citenamefont
  {Ivanov} \emph {et~al.}}]{torlina2015interpreting}%
  \BibitemOpen
  \bibfield  {author} {\bibinfo {author} {\bibfnamefont {L.}~\bibnamefont
  {Torlina}}, \bibinfo {author} {\bibfnamefont {F.}~\bibnamefont {Morales}},
  \bibinfo {author} {\bibfnamefont {J.}~\bibnamefont {Kaushal}}, \bibinfo
  {author} {\bibfnamefont {I.}~\bibnamefont {Ivanov}}, \bibinfo {author}
  {\bibfnamefont {A.}~\bibnamefont {Kheifets}}, \bibinfo {author}
  {\bibfnamefont {A.}~\bibnamefont {Zielinski}}, \bibinfo {author}
  {\bibfnamefont {A.}~\bibnamefont {Scrinzi}}, \bibinfo {author} {\bibfnamefont
  {H.~G.}\ \bibnamefont {Muller}}, \bibinfo {author} {\bibfnamefont
  {S.}~\bibnamefont {Sukiasyan}}, \bibinfo {author} {\bibfnamefont
  {M.}~\bibnamefont {Ivanov}}, \emph {et~al.},\ }\href@noop {} {\bibfield
  {journal} {\bibinfo  {journal} {Nature Physics}\ }\textbf {\bibinfo {volume}
  {11}},\ \bibinfo {pages} {503} (\bibinfo {year} {2015})}\BibitemShut
  {NoStop}%
\bibitem [{\citenamefont {Eckle}\ \emph
  {et~al.}(2008{\natexlab{a}})\citenamefont {Eckle}, \citenamefont {Smolarski},
  \citenamefont {Schlup}, \citenamefont {Biegert}, \citenamefont {Staudte},
  \citenamefont {Sch{\"o}ffler}, \citenamefont {Muller}, \citenamefont
  {D{\"o}rner},\ and\ \citenamefont {Keller}}]{eckle2008attosecond}%
  \BibitemOpen
  \bibfield  {author} {\bibinfo {author} {\bibfnamefont {P.}~\bibnamefont
  {Eckle}}, \bibinfo {author} {\bibfnamefont {M.}~\bibnamefont {Smolarski}},
  \bibinfo {author} {\bibfnamefont {P.}~\bibnamefont {Schlup}}, \bibinfo
  {author} {\bibfnamefont {J.}~\bibnamefont {Biegert}}, \bibinfo {author}
  {\bibfnamefont {A.}~\bibnamefont {Staudte}}, \bibinfo {author} {\bibfnamefont
  {M.}~\bibnamefont {Sch{\"o}ffler}}, \bibinfo {author} {\bibfnamefont {H.~G.}\
  \bibnamefont {Muller}}, \bibinfo {author} {\bibfnamefont {R.}~\bibnamefont
  {D{\"o}rner}},\ and\ \bibinfo {author} {\bibfnamefont {U.}~\bibnamefont
  {Keller}},\ }\href {https://www.nature.com/articles/nphys982} {\bibfield
  {journal} {\bibinfo  {journal} {Nature Physics}\ }\textbf {\bibinfo {volume}
  {4}},\ \bibinfo {pages} {565} (\bibinfo {year}
  {2008}{\natexlab{a}})}\BibitemShut {NoStop}%
\bibitem [{\citenamefont {Eckle}\ \emph
  {et~al.}(2008{\natexlab{b}})\citenamefont {Eckle}, \citenamefont {Pfeiffer},
  \citenamefont {Cirelli}, \citenamefont {Staudte}, \citenamefont {Dorner},
  \citenamefont {Muller}, \citenamefont {Buttiker},\ and\ \citenamefont
  {Keller}}]{eckle2008attosecond2}%
  \BibitemOpen
  \bibfield  {author} {\bibinfo {author} {\bibfnamefont {P.}~\bibnamefont
  {Eckle}}, \bibinfo {author} {\bibfnamefont {A.}~\bibnamefont {Pfeiffer}},
  \bibinfo {author} {\bibfnamefont {C.}~\bibnamefont {Cirelli}}, \bibinfo
  {author} {\bibfnamefont {A.}~\bibnamefont {Staudte}}, \bibinfo {author}
  {\bibfnamefont {R.}~\bibnamefont {Dorner}}, \bibinfo {author} {\bibfnamefont
  {H.}~\bibnamefont {Muller}}, \bibinfo {author} {\bibfnamefont
  {M.}~\bibnamefont {Buttiker}},\ and\ \bibinfo {author} {\bibfnamefont
  {U.}~\bibnamefont {Keller}},\ }\href
  {https://www.science.org/doi/full/10.1126/science.1163439?casa_token=4om5oufwkNMAAAAA\%3AyLzio5BAYr4CW0xpAbAEKzxTNXK2S_ojPxudKgfLEb0uNDrIqmK5ivOya3utFpEKRVs_a-zvxpJfGA}
  {\bibfield  {journal} {\bibinfo  {journal} {science}\ }\textbf {\bibinfo
  {volume} {322}},\ \bibinfo {pages} {1525} (\bibinfo {year}
  {2008}{\natexlab{b}})}\BibitemShut {NoStop}%
\bibitem [{\citenamefont {Pfeiffer}\ \emph {et~al.}(2012)\citenamefont
  {Pfeiffer}, \citenamefont {Cirelli}, \citenamefont {Smolarski}, \citenamefont
  {Dimitrovski}, \citenamefont {Abu-Samha}, \citenamefont {Madsen},\ and\
  \citenamefont {Keller}}]{pfeiffer2012attoclock}%
  \BibitemOpen
  \bibfield  {author} {\bibinfo {author} {\bibfnamefont {A.~N.}\ \bibnamefont
  {Pfeiffer}}, \bibinfo {author} {\bibfnamefont {C.}~\bibnamefont {Cirelli}},
  \bibinfo {author} {\bibfnamefont {M.}~\bibnamefont {Smolarski}}, \bibinfo
  {author} {\bibfnamefont {D.}~\bibnamefont {Dimitrovski}}, \bibinfo {author}
  {\bibfnamefont {M.}~\bibnamefont {Abu-Samha}}, \bibinfo {author}
  {\bibfnamefont {L.~B.}\ \bibnamefont {Madsen}},\ and\ \bibinfo {author}
  {\bibfnamefont {U.}~\bibnamefont {Keller}},\ }\href
  {https://www.nature.com/articles/nphys2125} {\bibfield  {journal} {\bibinfo
  {journal} {Nature Physics}\ }\textbf {\bibinfo {volume} {8}},\ \bibinfo
  {pages} {76} (\bibinfo {year} {2012})}\BibitemShut {NoStop}%
\bibitem [{\citenamefont {Pfeiffer}\ \emph {et~al.}(2013)\citenamefont
  {Pfeiffer}, \citenamefont {Cirelli}, \citenamefont {Smolarski},\ and\
  \citenamefont {Keller}}]{pfeiffer2013recent}%
  \BibitemOpen
  \bibfield  {author} {\bibinfo {author} {\bibfnamefont {A.~N.}\ \bibnamefont
  {Pfeiffer}}, \bibinfo {author} {\bibfnamefont {C.}~\bibnamefont {Cirelli}},
  \bibinfo {author} {\bibfnamefont {M.}~\bibnamefont {Smolarski}},\ and\
  \bibinfo {author} {\bibfnamefont {U.}~\bibnamefont {Keller}},\ }\href
  {https://www.sciencedirect.com/science/article/pii/S0301010412000626?casa_token=7l0iNa0xO9wAAAAA:T3_ZKDhPSzbTPQAo5xq_kaWdPsI7sjCY7ZCfjgx2jPqI1iF1Ru1YPY0mKC8Hi5JSoiWD9Ud14g}
  {\bibfield  {journal} {\bibinfo  {journal} {Chemical Physics}\ }\textbf
  {\bibinfo {volume} {414}},\ \bibinfo {pages} {84} (\bibinfo {year}
  {2013})}\BibitemShut {NoStop}%
\bibitem [{\citenamefont {Sainadh}\ \emph {et~al.}(2019)\citenamefont
  {Sainadh}, \citenamefont {Xu}, \citenamefont {Wang}, \citenamefont
  {Atia-Tul-Noor}, \citenamefont {Wallace}, \citenamefont {Douguet},
  \citenamefont {Bray}, \citenamefont {Ivanov}, \citenamefont {Bartschat},
  \citenamefont {Kheifets} \emph {et~al.}}]{sainadh2019attosecond}%
  \BibitemOpen
  \bibfield  {author} {\bibinfo {author} {\bibfnamefont {U.~S.}\ \bibnamefont
  {Sainadh}}, \bibinfo {author} {\bibfnamefont {H.}~\bibnamefont {Xu}},
  \bibinfo {author} {\bibfnamefont {X.}~\bibnamefont {Wang}}, \bibinfo {author}
  {\bibfnamefont {A.}~\bibnamefont {Atia-Tul-Noor}}, \bibinfo {author}
  {\bibfnamefont {W.~C.}\ \bibnamefont {Wallace}}, \bibinfo {author}
  {\bibfnamefont {N.}~\bibnamefont {Douguet}}, \bibinfo {author} {\bibfnamefont
  {A.}~\bibnamefont {Bray}}, \bibinfo {author} {\bibfnamefont {I.}~\bibnamefont
  {Ivanov}}, \bibinfo {author} {\bibfnamefont {K.}~\bibnamefont {Bartschat}},
  \bibinfo {author} {\bibfnamefont {A.}~\bibnamefont {Kheifets}}, \emph
  {et~al.},\ }\href {https://www.nature.com/articles/s41586-019-1028-3}
  {\bibfield  {journal} {\bibinfo  {journal} {Nature}\ }\textbf {\bibinfo
  {volume} {568}},\ \bibinfo {pages} {75} (\bibinfo {year} {2019})}\BibitemShut
  {NoStop}%
\bibitem [{\citenamefont {Landsman}\ \emph {et~al.}(2014)\citenamefont
  {Landsman}, \citenamefont {Weger}, \citenamefont {Maurer}, \citenamefont
  {Boge}, \citenamefont {Ludwig}, \citenamefont {Heuser}, \citenamefont
  {Cirelli}, \citenamefont {Gallmann},\ and\ \citenamefont
  {Keller}}]{landsman2014ultrafast}%
  \BibitemOpen
  \bibfield  {author} {\bibinfo {author} {\bibfnamefont {A.~S.}\ \bibnamefont
  {Landsman}}, \bibinfo {author} {\bibfnamefont {M.}~\bibnamefont {Weger}},
  \bibinfo {author} {\bibfnamefont {J.}~\bibnamefont {Maurer}}, \bibinfo
  {author} {\bibfnamefont {R.}~\bibnamefont {Boge}}, \bibinfo {author}
  {\bibfnamefont {A.}~\bibnamefont {Ludwig}}, \bibinfo {author} {\bibfnamefont
  {S.}~\bibnamefont {Heuser}}, \bibinfo {author} {\bibfnamefont
  {C.}~\bibnamefont {Cirelli}}, \bibinfo {author} {\bibfnamefont
  {L.}~\bibnamefont {Gallmann}},\ and\ \bibinfo {author} {\bibfnamefont
  {U.}~\bibnamefont {Keller}},\ }\href
  {https://opg.optica.org/optica/fulltext.cfm?uri=optica-1-5-343&id=304679}
  {\bibfield  {journal} {\bibinfo  {journal} {Optica}\ }\textbf {\bibinfo
  {volume} {1}},\ \bibinfo {pages} {343} (\bibinfo {year} {2014})}\BibitemShut
  {NoStop}%
\bibitem [{\citenamefont {Camus}\ \emph {et~al.}(2017)\citenamefont {Camus},
  \citenamefont {Yakaboylu}, \citenamefont {Fechner}, \citenamefont {Klaiber},
  \citenamefont {Laux}, \citenamefont {Mi}, \citenamefont {Hatsagortsyan},
  \citenamefont {Pfeifer}, \citenamefont {Keitel},\ and\ \citenamefont
  {Moshammer}}]{camus2017experimental}%
  \BibitemOpen
  \bibfield  {author} {\bibinfo {author} {\bibfnamefont {N.}~\bibnamefont
  {Camus}}, \bibinfo {author} {\bibfnamefont {E.}~\bibnamefont {Yakaboylu}},
  \bibinfo {author} {\bibfnamefont {L.}~\bibnamefont {Fechner}}, \bibinfo
  {author} {\bibfnamefont {M.}~\bibnamefont {Klaiber}}, \bibinfo {author}
  {\bibfnamefont {M.}~\bibnamefont {Laux}}, \bibinfo {author} {\bibfnamefont
  {Y.}~\bibnamefont {Mi}}, \bibinfo {author} {\bibfnamefont {K.~Z.}\
  \bibnamefont {Hatsagortsyan}}, \bibinfo {author} {\bibfnamefont
  {T.}~\bibnamefont {Pfeifer}}, \bibinfo {author} {\bibfnamefont {C.~H.}\
  \bibnamefont {Keitel}},\ and\ \bibinfo {author} {\bibfnamefont
  {R.}~\bibnamefont {Moshammer}},\ }\href
  {https://journals.aps.org/prl/abstract/10.1103/PhysRevLett.119.023201}
  {\bibfield  {journal} {\bibinfo  {journal} {Physical review letters}\
  }\textbf {\bibinfo {volume} {119}},\ \bibinfo {pages} {023201} (\bibinfo
  {year} {2017})}\BibitemShut {NoStop}%
\bibitem [{\citenamefont {Galapon}\ and\ \citenamefont
  {Magadan}(2018)}]{galapon2018quantizations}%
  \BibitemOpen
  \bibfield  {author} {\bibinfo {author} {\bibfnamefont {E.~A.}\ \bibnamefont
  {Galapon}}\ and\ \bibinfo {author} {\bibfnamefont {J.~J.~P.}\ \bibnamefont
  {Magadan}},\ }\href@noop {} {\bibfield  {journal} {\bibinfo  {journal}
  {Annals of Physics}\ }\textbf {\bibinfo {volume} {397}},\ \bibinfo {pages}
  {278} (\bibinfo {year} {2018})}\BibitemShut {NoStop}%
\bibitem [{\citenamefont {Sombillo}\ and\ \citenamefont
  {Galapon}(2014)}]{sombillo2014quantum}%
  \BibitemOpen
  \bibfield  {author} {\bibinfo {author} {\bibfnamefont {D.~L.}\ \bibnamefont
  {Sombillo}}\ and\ \bibinfo {author} {\bibfnamefont {E.~A.}\ \bibnamefont
  {Galapon}},\ }\href@noop {} {\bibfield  {journal} {\bibinfo  {journal}
  {Physical Review A}\ }\textbf {\bibinfo {volume} {90}},\ \bibinfo {pages}
  {032115} (\bibinfo {year} {2014})}\BibitemShut {NoStop}%
\bibitem [{\citenamefont {Galapon}(2012)}]{PhysRevLett.108.170402}%
  \BibitemOpen
  \bibfield  {author} {\bibinfo {author} {\bibfnamefont {E.~A.}\ \bibnamefont
  {Galapon}},\ }\href {https://doi.org/10.1103/PhysRevLett.108.170402}
  {\bibfield  {journal} {\bibinfo  {journal} {Phys. Rev. Lett.}\ }\textbf
  {\bibinfo {volume} {108}},\ \bibinfo {pages} {170402} (\bibinfo {year}
  {2012})}\BibitemShut {NoStop}%
\bibitem [{\citenamefont {Pablico}\ and\ \citenamefont
  {Galapon}(2020)}]{pablico2020quantum}%
  \BibitemOpen
  \bibfield  {author} {\bibinfo {author} {\bibfnamefont {D.~A.~L.}\
  \bibnamefont {Pablico}}\ and\ \bibinfo {author} {\bibfnamefont {E.~A.}\
  \bibnamefont {Galapon}},\ }\href@noop {} {\bibfield  {journal} {\bibinfo
  {journal} {Physical Review A}\ }\textbf {\bibinfo {volume} {101}},\ \bibinfo
  {pages} {022103} (\bibinfo {year} {2020})}\BibitemShut {NoStop}%
\bibitem [{\citenamefont {Flores}\ and\ \citenamefont
  {Galapon}(2023{\natexlab{a}})}]{flores2023instantaneous}%
  \BibitemOpen
  \bibfield  {author} {\bibinfo {author} {\bibfnamefont {P.~C.}\ \bibnamefont
  {Flores}}\ and\ \bibinfo {author} {\bibfnamefont {E.~A.}\ \bibnamefont
  {Galapon}},\ }\href@noop {} {\bibfield  {journal} {\bibinfo  {journal}
  {Europhysics Letters}\ }\textbf {\bibinfo {volume} {141}},\ \bibinfo {pages}
  {10001} (\bibinfo {year} {2023}{\natexlab{a}})}\BibitemShut {NoStop}%
\bibitem [{\citenamefont {Flores}\ and\ \citenamefont
  {Galapon}(2023{\natexlab{b}})}]{flores2023quantized}%
  \BibitemOpen
  \bibfield  {author} {\bibinfo {author} {\bibfnamefont {P.}~\bibnamefont
  {Flores}}\ and\ \bibinfo {author} {\bibfnamefont {E.~A.}\ \bibnamefont
  {Galapon}},\ }\href@noop {} {\bibfield  {journal} {\bibinfo  {journal} {The
  European Physical Journal Plus}\ }\textbf {\bibinfo {volume} {138}},\
  \bibinfo {pages} {1} (\bibinfo {year} {2023}{\natexlab{b}})}\BibitemShut
  {NoStop}%
\bibitem [{\citenamefont {Steinberg}\ \emph {et~al.}(1993)\citenamefont
  {Steinberg}, \citenamefont {Kwiat},\ and\ \citenamefont
  {Chiao}}]{steinberg1993measurement}%
  \BibitemOpen
  \bibfield  {author} {\bibinfo {author} {\bibfnamefont {A.~M.}\ \bibnamefont
  {Steinberg}}, \bibinfo {author} {\bibfnamefont {P.~G.}\ \bibnamefont
  {Kwiat}},\ and\ \bibinfo {author} {\bibfnamefont {R.~Y.}\ \bibnamefont
  {Chiao}},\ }\href
  {https://journals.aps.org/prl/abstract/10.1103/PhysRevLett.71.708} {\bibfield
   {journal} {\bibinfo  {journal} {Physical Review Letters}\ }\textbf {\bibinfo
  {volume} {71}},\ \bibinfo {pages} {708} (\bibinfo {year} {1993})}\BibitemShut
  {NoStop}%
\bibitem [{\citenamefont {Aharonov}\ and\ \citenamefont
  {Bohm}(1961)}]{aharonov1961time}%
  \BibitemOpen
  \bibfield  {author} {\bibinfo {author} {\bibfnamefont {Y.}~\bibnamefont
  {Aharonov}}\ and\ \bibinfo {author} {\bibfnamefont {D.}~\bibnamefont
  {Bohm}},\ }\href@noop {} {\bibfield  {journal} {\bibinfo  {journal} {Physical
  Review}\ }\textbf {\bibinfo {volume} {122}},\ \bibinfo {pages} {1649}
  (\bibinfo {year} {1961})}\BibitemShut {NoStop}%
\bibitem [{\citenamefont {Galapon}(2004)}]{galapon2004shouldn}%
  \BibitemOpen
  \bibfield  {author} {\bibinfo {author} {\bibfnamefont {E.~A.}\ \bibnamefont
  {Galapon}},\ }\href@noop {} {\bibfield  {journal} {\bibinfo  {journal}
  {Journal of mathematical physics}\ }\textbf {\bibinfo {volume} {45}},\
  \bibinfo {pages} {3180} (\bibinfo {year} {2004})}\BibitemShut {NoStop}%
\bibitem [{\citenamefont {Kullie}(2020)}]{Kullie2020}%
  \BibitemOpen
  \bibfield  {author} {\bibinfo {author} {\bibfnamefont {O.}~\bibnamefont
  {Kullie}},\ }\href@noop {} {\bibfield  {journal} {\bibinfo  {journal}
  {Quantum Rep.}\ }\textbf {\bibinfo {volume} {2}},\ \bibinfo {pages} {233}
  (\bibinfo {year} {2020})}\BibitemShut {NoStop}%
\bibitem [{\citenamefont {Sainadh}\ \emph {et~al.}(2020)\citenamefont
  {Sainadh}, \citenamefont {Sang},\ and\ \citenamefont
  {Litvinyuk}}]{Sainadh2020}%
  \BibitemOpen
  \bibfield  {author} {\bibinfo {author} {\bibfnamefont {U.}~\bibnamefont
  {Sainadh}}, \bibinfo {author} {\bibfnamefont {R.}~\bibnamefont {Sang}},\ and\
  \bibinfo {author} {\bibfnamefont {I.}~\bibnamefont {Litvinyuk}},\ }\href@noop
  {} {\bibfield  {journal} {\bibinfo  {journal} {JPhys Photonics}\ }\textbf
  {\bibinfo {volume} {2}},\ \bibinfo {pages} {042002} (\bibinfo {year}
  {2020})}\BibitemShut {NoStop}%
\bibitem [{\citenamefont {Kullie}(2015)}]{Kullie2015}%
  \BibitemOpen
  \bibfield  {author} {\bibinfo {author} {\bibfnamefont {O.}~\bibnamefont
  {Kullie}},\ }\href@noop {} {\bibfield  {journal} {\bibinfo  {journal} {Phys.
  Rev. A}\ }\textbf {\bibinfo {volume} {92}},\ \bibinfo {pages} {052118}
  (\bibinfo {year} {2015})}\BibitemShut {NoStop}%
\end{thebibliography}%

\end{document}